\documentclass[aps,pra,reprint,showpacs,superscriptaddress,twocolumn]{revtex4-1}
\usepackage{amsmath}
\usepackage{array}
\usepackage{graphicx}
\usepackage{amssymb}
\usepackage{amsthm}
\usepackage{txfonts}
\usepackage[colorlinks,linkcolor=blue,citecolor=blue,hyperindex]{hyperref}
\begin{document}

\title{Correlation Measurement of an unknown state with Weak Coupling}
\author{Hui Li}
\affiliation{School of Science, Tianjin University of Technology, Tianjin 300384,
China}
\email[]{lihuiwuli@126.com}

\author{Yan-Song Li}
\email[]{ysli@mail.tsinghua.edu.cn}
\affiliation{State Key Laboratory of Low-Dimensional Quantum Physics and Department of Physics, Tsinghua University, Beijing 100084,
China}
\affiliation{Tsinghua National Laboratory for Information Science
and Technology, Beijing 100084, China}

\author{Chao Zheng}
\affiliation{Department of Physics, College of Science, North China University of Technology, Beijing 100144, China}

\author{Xian Lu}
\affiliation{Technology on Integrated Information System Laboratory, Institute of Software, CAS, Beijing 100190, China}

\date{\today }

\begin{abstract}

Traditionally, quantum state correlation can be obtained with  calculations on a state density matrix already known. Here, we propose a model with which correlations of unknown quantum states can be obtained. There are no needs of classical communication in the course of coupling, optimization and complicated calculations. All we need are weak coupling and ancillary systems. We detail the model on the state in which particles belong to the different owners. A concisely example is elaborated in the last part of this paper.

\end{abstract}


\maketitle

\section{Introduction}
Correlation plays a significant important role in quantum calculation and communication \cite{book1,prl-69-2881,prl-70-1992,prl-67-661}. Studying the measure of quantum correlations has been an active research area for over a long time and has been given rise to various correlations measure method. For example: Entanglement, regarded as the quantum mechanical property, is well defined  for pure state \cite{rmp-81-865}, but there exist several definitions when it comes to mixed states. Entanglement of Formation \cite{prl-80-2245,prl-95-210501}, Computable measure of entanglement \cite{pra-65-032314}, Logarithmic Nagativity \cite{prl-95-090503}, concurrence \cite{prl-78-5022}. Quantum discord,  based on von Neumann measurements and majorization, is another calculation method proposed in recent years set up state correlations at a certain level \cite{prl-88-017901,jpa-34-6899}.

Previously, general correlations calculation methods have been put forward to work in the circumstances of  we already know the quantum state density matrix. This calculation work was performed through the matrix elements. In this paper, we describe a general model that allows for getting correlations of unknown quantum states. Here, we make use of the weak measurement described in Ref. \cite{prl-60-1351,book2} and the ancillary parts described in Ref. \cite{prl-116-070404}. The unknown state particles respectively belong to different owners, and all of these owners know nothing about the information of their particles but the dimension. After series of operations according to the model are complete, one of these owners can obtain correlations of the whole system consisted by the particles. The operations are executed by each owner individually, and there is no classical information needed during the whole thing happens.

Our paper is organized as follows. In Sec. \ref{sec:2} we describe our model to obtain quantum correlations of unknown quantum states. It is divided into three subsections. In the first subsection, we explain weak measurement and its application shown in Ref. \cite{scir-3-1193}. In the second subsection, we elaborate our model. Every step of the model is illustrated at great length. In the third subsection, we present some details about our model. We further note that these details are the key points of realization of our model. In Sec. \ref{sec:3} we give an example of three particles system where the model proposed in this paper is used. A discussion and a summary are given in Sec. \ref{sec:4}.

\section{THEORETICAL MODEL}
\label{sec:2}

\subsection{Weak measurement and its application}
The concept of weak values in quantum mechanics was formulated by Aharonov, Albert, and Vaidman \cite{prl-60-1351,book2}. We consider an initial state $\left| {{\psi _{in}}} \right\rangle  $ and a final state $\left| {{\psi _{fin}}} \right\rangle $. A measurement
interaction is ${H_{{\mathop{\rm int}} }}\left( t \right) = g(t){\kern 1pt} {A}{\kern 1pt} {P_d}$. The coupling $g(t)$ is non zero for times $0\le t\le T$  and normalized as $\int_0^T {g\left( t \right)dt}  = {g_0}$.
Then, the weak value of $A$ is
\begin{eqnarray}
{\left\langle A \right\rangle _w} = \frac{{\left\langle {{\psi _{fin}}} \right.\left| A \right|\left. {{\psi _{in}}} \right\rangle }}{{\langle {\psi _{fin}}|\left. {{\psi _{in}}} \right\rangle }},
\label{eq1}
\end{eqnarray}
where $A$ is the measured observable operator, and ${P_d}$ is conjugate to an observable ${Q_d}$ presenting the pointer position on the measuring device.

There has been significant interests in the applications of weak value since it was presented \cite{prl-108-070402,pra-65-032111}, including the one that was described in Ref. \cite{scir-3-1193} which be used in our model. We consider an unknown mixed state $\rho $ and its correlation  to be obtained, a complete set of projective operators ${A_i} = \left| {{a_i}} \right\rangle \left\langle {{a_i}} \right|$ used to perform weak measurements, and a final projective measurement basis set $\left| {{b_k}} \right\rangle $ acting as postselected states. The matrix elements of  $\rho $ can be expressed as
\begin{eqnarray}
 \left\langle {{a_i}} \right.\left| \rho  \right|\left. {{a_j}} \right\rangle  = \sum\limits_k {\left\langle {{b_k}} \right.\left| \rho  \right|\left. {{b_k}} \right\rangle \frac{{\left\langle {{{b_k}}}
 \mathrel{\left | {\vphantom {{{b_k}} {{a_j}}}}
 \right. \kern-\nulldelimiterspace}
 {{{a_j}}} \right\rangle }}{{\left\langle {{{b_k}}}
 \mathrel{\left | {\vphantom {{{b_k}} {{a_i}}}}
 \right. \kern-\nulldelimiterspace}
 {{{a_i}}} \right\rangle }}{W_{ki}}}  = \sum\limits_k {{P_k}\frac{{{\beta _{kj}}}}{{{\beta _{ki}}}}{W_{ki}}}.\nonumber \\
\label{eq2}
\end{eqnarray}
where ${{W_{ki}}}$ is the weak value of projective operator $\left| {{a_i}} \right\rangle \left\langle {{a_i}} \right|$ with the system state being postselected by basis $\left| {{b_k}} \right\rangle \left\langle {{b_k}} \right|$.

\subsection{Correlation measurement of a unknown quantum state}
\begin{figure}[!htb]
\begin{center}
\includegraphics[width=8.5cm,angle=0]{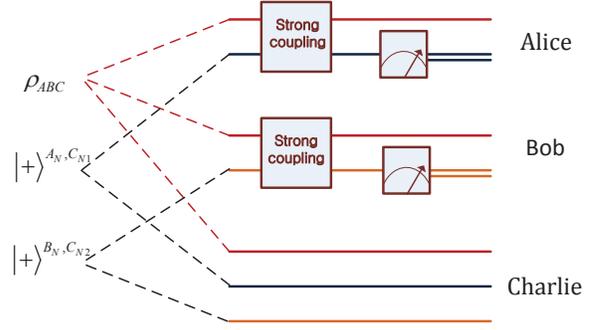}
\caption{(Color online)  Local operations executed by  Alice and Bob. Strong couplings and measurements are locally executed, conveying the state composed of system  $A$, $B$ and $C$ to the system composed of $C$, ${C_{N1}}$ and  ${C_{N2}}$. }
\label{FIG1}
\end{center}
\end{figure}

Suppose a density matrix of quantum systems $A$ , $B$ and $C$ is $\rho  = \sum\limits_n {{p_n}\left| {{\psi _n}} \right\rangle \langle {\psi _n}|}  $, where ${\left| {{\psi _i}} \right\rangle }$ is the pure-state decompositions member. It is convenient to rewrite the density matrix in the form $\rho  = \sum\limits_{ijk} {{\alpha _{ijk}}\left| {i,j,k} \right\rangle \left\langle {i,j,k} \right|} $. ${N_1}$ and ${N_2}$ are auxiliary systems which will be prepared in the initial entangled states 
${\left|  +  \right\rangle ^{{A_N},{C_{{N_1}}}}} = \left( {{1 \mathord{\left/
 {\vphantom {1 {\sqrt l }}} \right.
 \kern-\nulldelimiterspace} {\sqrt {l_1} }}} \right)\sum\limits_m {{{\left| m \right\rangle }^{{A_N}}}{{\left| m \right\rangle }^{{C_{{N_1}}}}}}  $ 
and 
${\left|  +  \right\rangle ^{{B_N},{C_{{N_2}}}}} = \left( {{1 \mathord{\left/
 {\vphantom {1 {\sqrt {{l_2}} }}} \right.
 \kern-\nulldelimiterspace} {\sqrt {l_2} }}} \right)\sum\limits_m {{{\left| m \right\rangle }^{{B_N}}}{{\left| m \right\rangle }^{{C_{N2}}}}}  $. 
Where ${{l_1}}$ and ${{l_2}}$ are the dimensions of particles  $A$ and $B$, which may have the same value or not.
Each of the quantum system's particles respectively belongs to three different owners: Alice, Bob and Charlie. The three owners also respectively have the auxiliary particles: ${A_N}$ belongs to Alice, ${B_N}$ belongs to Bob, ${C_{N1}}$ and ${C_{N2}}$ belong to Charlie.
In the area controlled by Alice and Bob, the system particles are coupled to the auxiliary particles by the owners they belong to, then the coupling  is followed by a local measurement on the auxiliary: the particle $A$ is strongly coupled to particle ${A_N}$ in the area controlled by Alice, and $B$ is coupled to ${B_N}$. Alice and Bob respectively measure the particle ${A_N}$ and  ${B_N}$ after the couplings are executed. The particles without being coupled and measured: $C$, ${C_{N1}}$ and ${C_{N2}}$ are all under control of Charlie (see Fig. \ref{FIG1}). The function of the operations above all is conveying the state of system $A$, $B$ and $C$ to the system of $C$, ${C_{N1}}$ and  ${C_{N2}}$, that is ${\rho _{{C_{N1}}{C_{N2}}C}}$. The details of couplings and measurings were set forth in Ref. \cite{prl-116-070404}.

\begin{figure}[!htb]
\begin{center}
\includegraphics[width=8.5cm,angle=0]{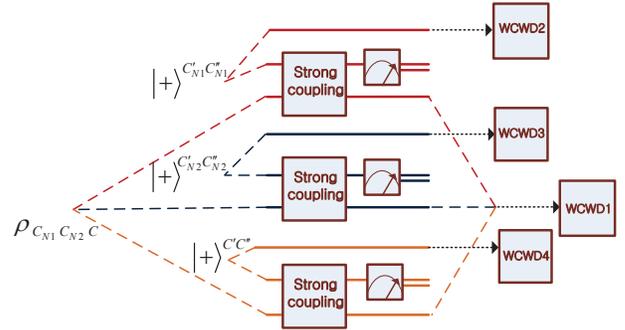}
\caption{(Color online)  Local operations executed by Charlie. ${C_{N1}}$, ${C_{N2}}$ and $C$ are respectively strong coupled to auxiliary particles ${{C'}_{N1}}$, ${{C'}_{N2}}$ and ${C'}$. There are four lines to be continued including three auxiliary lines and ${\rho _{{C_{N1}}{C_{N2}}C}}$ line: WCWD1,WCWD2,WCWD3 and WCWD4.  }
\label{FIG2}
\end{center}
\end{figure}

\begin{figure}[!htb]
\begin{center}
\includegraphics[width=8.5cm,angle=0]{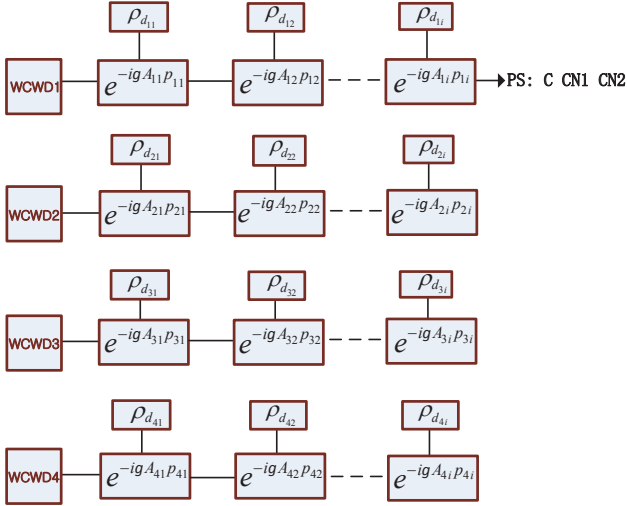}
\caption{ Four lines of WCWD. The functions of  WCWD1 is to abstract the information of state ${\rho _{{C_{N1}}{C_{N2}}C}}$,  and the functions of WCWD2, WCWD3, WCWD4 is to respectively abstract the information of  ${\rho _A}$,  ${\rho _B}$ and  ${\rho _C}$. $\mathtt{g}$ is the coupling strength parameter of the weak measurement. }
\label{FIG3}
\end{center}
\end{figure}

Here, Charlie is the one who is going to obtain the correlations of the system ${\rho _{ABC}}$, and he continue to operate the particles ${C_{N1}}$, ${C_{N2}}$ and $C$ (see Fig. \ref{FIG2}). ${\left|  +  \right\rangle ^{{{C'}_{N1}}{{C''}_{N1}}}}$, ${\left|  +  \right\rangle ^{{{C'}_{N2}}{{C''}_{N2}}}}$ and ${\left|  +  \right\rangle ^{C'C''}}$ are auxiliary systems prepared in the entangled state which  ${C_{N1}}$, ${C_{N2}}$ and $C$ are respectively strong coupled to. There are four lines of  WCWD (weak coupling work device) to be continued to work on including three auxiliary lines and ${\rho _{{C_{N1}}{C_{N2}}C}}$ line  : WCWD1, WCWD2, WCWD3 and WCWD4. The continued work is shown in Fig. \ref{FIG3}.

According to  Fig.\ref{FIG3}, all weak couplings  are connected with a measuring device presenting the point position shift and momentum shift which give the weak value of measured operator through some calculations \cite{pra-76-044103}. We can consider all measuring device in Fig.\ref{FIG3} as a matrix.  In the WCWD1 $\left\{ {{a_{11}},{a_{12}}, \cdots {a_{1i}}} \right\}$ is the base set of system ${\rho _{{C_{N1}}{C_{N2}}C}}$, and $\left\{ {{A_{11}},{A_{12}}, \cdots {A_{1i}}} \right\} = \left\{ {\left| {{a_{11}}} \right\rangle \left\langle {{a_{11}}} \right|,\left| {{a_{12}}} \right\rangle \left\langle {{a_{12}}} \right|, \cdots \left| {{a_{1i}}} \right\rangle \left\langle {{a_{1i}}} \right|} \right\}$.  ${a_{2i}}$ in the WCWD2 is the element of the base set of system ${\rho _A}$ picked out from ${a_{1i}}$ in the same column of weak couplings device matrix, and we arrange ${a_{3i}}$ and ${a_{4i}}$ the same way which belong to the base set of ${\rho _B}$ and ${\rho _C}$. After all these coupling work is done, a postselected  operation on $C$, ${C_{N1}}$ and ${C_{N2}}$ is operated by Charlie to induce all meters point to present shifting.
\begin{eqnarray}
{W_{ki}} = \frac{{{\rm{tr}}\left\{ {\left( {\left| {{b_k}} \right\rangle \langle {b_k}|} \right)\left( {\left| {{a_{ji}}} \right\rangle \langle {a_{ji}}|} \right)\rho } \right\}}}{{{\rm{tr}}\left\{ {\left( {\left| {{b_k}} \right\rangle \langle {b_k}|} \right)\rho } \right\}}}
\label{eq3}
\end{eqnarray}
where if $j = 1$ the information of the whole system are abstracted, and the subsystem's information is abstracted if $j = 2,3,4$.  $\left\{ {{b_1},{b_2}, \cdots {b_k}} \right\}$ is a baseset that is mutually unbiased to  $\left\{ {{a_{11}},{a_{12}}, \cdots {a_{1i}}} \right\}$. The correlation of ${\rho _{ABC}}$ is given by 
\begin{eqnarray}
{C_{{\rho _{ABC}}}} = \sum\limits_{k = 1}^n {{P_k}\left\{ {\sum\limits_{i = 1}^n {\left| {\varepsilon \left( {{\rho _{{d_{1i}}}}} \right) - \prod\limits_{j = 2}^4 {\varepsilon \left( {{\rho _{{d_{ji}}}}} \right)} } \right|} } \right\}} 
\label{eq4}
\end{eqnarray}
where, $\varepsilon $ is a process that obtain weak value from the point shift of device ${\rho _{{d_{ji}}}}$: ${\rho _{{d_{ji}}}} \to {W_{ji}}$.

\subsection{Some details}
The model in this paper exploit the concept of trace distance explained in Ref.\cite{book1}, and the distance between the whole system ${\rho _{ABC}}$ and the separable system ${\rho _A} \otimes {\rho _B} \otimes {\rho _C}$ can be regarded as correlations of ${\rho _{ABC}}$ at a certain level. If the two states are commute they are diagonal in the same basis, then the trace distance is the sum of the variations of density matrix diagonal elements  ${D_{(\rho ,\sigma )}} = \frac{1}{2}{\rm tr}\left| {\sum\limits_i {\left( {{r_i} - {s_i}} \right)\left| i \right\rangle \left\langle i \right|} } \right|$, which is the theoretical support for our model and manipulation Eq. \ref{eq4}.

When the postseletion  take place, the shift of different pointers of  four lines device in Fig. \ref{FIG3}, including the position shift and the momentum shift, can be read out simultaneously .
It can be neglected that the chance of introducing state variation by weak measurement if the number of measurement device is not very large. But if the number of measurement device is very large or the coupling strength parameter $\mathtt{g}$ is not sufficiently small, we have to take into account the influence state yielded by the large amount of weak measurement. To reduce this influence, we arrange three strong couplings (Fig.\ref{FIG2}) after the first strong coupling in Fig.\ref{FIG1} to outspread another three  lines: WCDW2, WCDW3 and WCDW4, which are in charge of the sigle particle weak measurement. If coupling strength parameter $\mathtt{g}$ is enough small or the dimension of the system (the number of weak measurement device) is not very large, we can skip over the operations in Fig.\ref{FIG2}  to Fig.\ref{FIG3}, and incorperate the weak measurement devices of  WCDW2, WCDW3 and WCDW4 into  WCDW1.

When it comes to the postselection, we do postselection in WCDW1 on $C$, ${C_{N1}}$ and ${C_{N2}}$ the system particles in WCDW2, WCDW3, and WCDW4 collapse to the postselected state in WCDW1 at the same time. The particles $A$ and $B$ in Fig. \ref{FIG1} are also collapse with the postselection being done in Fig. \ref{FIG3}. So the postselection in WCDW1 is the only postselected operation and is enough for the model in this paper. 

Suppose that $\mu $ is the value we get from strong measurement meters In Fig. \ref{FIG2}.  For $\mu  = 0$ this is the desired result, we have ${\rho _{{C_{N1}}}} \otimes {\rho _{{C_{N2}}}} \otimes {\rho _C}$ obtained directly. For $\mu  \ne 0$, we obtain the state that have the same spectrum as ${\rho _{{C_{N1}}}} \otimes {\rho _{{C_{N2}}}} \otimes {\rho _C}$. So the result of system state correlations cannot be changed with the variance of $\mu $ according to Eq. \ref{eq4}. But there is one concept we have to agree with, that the three strong measurement meters show the same reading. This can be easily realized by means of projecting ${{C''}_{N1}}$, ${{C''}_{N2}}$ and ${C''}$ on a same base. Suppose that ${\nu _1}$ and ${\nu _2}$ are the two strong meter readings in Fig.\ref{FIG1}. For ${\nu _1} = 0$ and ${\nu _2} = 0$  this is also the desired result. We get ${\rho _{{C_{N1}}{C_{N2}}C}} = {\rho _{ABC}}$ given this desired result. But if ${\nu _1} \ne 0$, ${\nu _2} \ne 0$ and even if ${\nu _1} \ne {\nu _2}$, ${\rho _{{C_{N1}}{C_{N2}}C}}$ is the local unitary transform of ${\rho _{ABC}}$, so the correlations of state ${\rho _{ABC}}$ can also be obtained through our model and Eq.\ref{eq4}.

\begin{table*}
\caption{\label{table1}.Weak measured operators in Fig. \ref{FIG2}. ${A_{1i}}$, ${A_{2i}}$, ${A_{3i}}$ and ${A_{4i}}$ are the weak measured observable operators. All operators are displayed in this table. $\left\{ {{A_{11}},{A_{12}}, \cdots {A_{1i}}} \right\}$ is the base set of system ${\rho _{{C_{N1}}{C_{N2}}C}}$. The  ${A_{2i}}$ is the element of the base set of system ${\rho _A}$ picked out from ${A_{1i}}$ that is in the same column of weak couplings device matrix, and we arrange ${A_{3i}}$ and ${A_{4i}}$ the same way which belong to the base set of ${\rho _B}$ and ${\rho _C}$. }
\begin{ruledtabular}
\begin{tabular}{ccccccccc}
$i$ &1&2&3&4&5&6&7&8\\ \hline
${A_{1i}}$&$\left| {000} \right\rangle \left\langle {000} \right|$&$\left| {001} \right\rangle \left\langle {001} \right|$&$\left| {010} \right\rangle \left\langle {010} \right|$&$\left| {011} \right\rangle \left\langle {011} \right|$&$\left| {100} \right\rangle \left\langle {100} \right|$&$\left| {101} \right\rangle \left\langle {101} \right|$&$\left| {110} \right\rangle \left\langle {110} \right|$&$\left| {111} \right\rangle \left\langle {111} \right|$\\
${A_{2i}}$&$\left| 0 \right\rangle \left\langle 0 \right|$&$\left| 0 \right\rangle \left\langle 0 \right|$&$\left| 0 \right\rangle \left\langle 0 \right|$&$\left| 0 \right\rangle \left\langle 0 \right|$&$\left| 1 \right\rangle \left\langle 1 \right|$&$\left| 1 \right\rangle \left\langle 1 \right|$&$\left| 1 \right\rangle \left\langle 1 \right|$&$\left| 1 \right\rangle \left\langle 1 \right|$\\
${A_{3i}}$&$\left| 0 \right\rangle \left\langle 0 \right|$&$\left| 0 \right\rangle \left\langle 0 \right|$&$\left| 1 \right\rangle \left\langle 1 \right|$&$\left| 1 \right\rangle \left\langle 1 \right|$&$\left| 0 \right\rangle \left\langle 0 \right|$&$\left| 0 \right\rangle \left\langle 0 \right|$&$\left| 1 \right\rangle \left\langle 1 \right|$&$\left| 1 \right\rangle \left\langle 1 \right|$\\
${A_{4i}}$&$\left| 0 \right\rangle \left\langle 0 \right|$&$\left| 1 \right\rangle \left\langle 1 \right|$&$\left| 0 \right\rangle \left\langle 0 \right|$&$\left| 1 \right\rangle \left\langle 1 \right|$&$\left| 0 \right\rangle \left\langle 0 \right|$&$\left| 1 \right\rangle \left\langle 1 \right|$&$\left| 0 \right\rangle \left\langle 0 \right|$&$\left| 1 \right\rangle \left\langle 1 \right|$\\
\end{tabular}
\end{ruledtabular}
\end{table*}

\begin{table}[b]
\caption{\label{table2} The postselected states. $\left\{ {{b_1},{b_2}, \cdots {b_k}} \right\}$ is a baseset that is mutually unbiased to  $\left\{ {{a_{11}},{a_{12}}, \cdots {a_{1i}}} \right\}$. $\left\{ {{a_{11}},{a_{12}}, \cdots {a_{1i}}} \right\}$ is the baseset of the system whose correlation we aim to obtain.}
\begin{ruledtabular}
\begin{tabular}{cc}
$k$&${b_k}$\\
\hline
1& $\left| {000} \right\rangle  + \left| {001} \right\rangle  + \left| {010} \right\rangle  + \left| {011} \right\rangle  + \left| {100} \right\rangle  + \left| {101} \right\rangle  + \left| {110} \right\rangle  + \left| {111} \right\rangle$\\
2&$\left| {000} \right\rangle  - \left| {001} \right\rangle  + \left| {010} \right\rangle  - \left| {011} \right\rangle  + \left| {100} \right\rangle  - \left| {101} \right\rangle  + \left| {110} \right\rangle  - \left| {111} \right\rangle $\\
3&$\left| {000} \right\rangle  + \left| {001} \right\rangle  - \left| {010} \right\rangle  - \left| {011} \right\rangle  + \left| {100} \right\rangle  + \left| {101} \right\rangle  - \left| {110} \right\rangle  - \left| {111} \right\rangle $\\
4&$\left| {000} \right\rangle  - \left| {001} \right\rangle  - \left| {010} \right\rangle  + \left| {011} \right\rangle  + \left| {100} \right\rangle  - \left| {101} \right\rangle  - \left| {110} \right\rangle  + \left| {111} \right\rangle $\\
5&$\left| {000} \right\rangle  + \left| {001} \right\rangle  + \left| {010} \right\rangle  + \left| {011} \right\rangle  - \left| {100} \right\rangle  - \left| {101} \right\rangle  - \left| {110} \right\rangle  - \left| {111} \right\rangle $\\
6&$\left| {000} \right\rangle  - \left| {001} \right\rangle  + \left| {010} \right\rangle  - \left| {011} \right\rangle  - \left| {100} \right\rangle  + \left| {101} \right\rangle  - \left| {110} \right\rangle  + \left| {111} \right\rangle $\\
7&$\left| {000} \right\rangle  + \left| {001} \right\rangle  - \left| {010} \right\rangle  - \left| {011} \right\rangle  - \left| {100} \right\rangle  - \left| {101} \right\rangle  + \left| {110} \right\rangle  + \left| {111} \right\rangle $\\
8&$\left| {000} \right\rangle  - \left| {001} \right\rangle  - \left| {010} \right\rangle  + \left| {011} \right\rangle  - \left| {100} \right\rangle  + \left| {101} \right\rangle  + \left| {110} \right\rangle  - \left| {111} \right\rangle $\\
\end{tabular}
\end{ruledtabular}
\end{table}

\section{EXAMPLE IMPLEMENTATIONS}
\label{sec:3}

We now consider a system ${\rho _{ABC}}$ with three two-dimensional particles as an illustration to be displayed in the following section. Alice, Bob and Charlie individually have particle $A$, $B$ and $C$ at disposal.  They do not have any idea of the information of this system except its dimensions and base set. Two bipartite maximal entangled states are prepared to be used as auxiliary systems and named ${\left|  +  \right\rangle ^{{A_N},{C_{N1}}}}$ and ${\left|  +  \right\rangle ^{{B_N},{C_{N2}}}}$. Alice possess the first particle of ${\left|  +  \right\rangle ^{{A_N},{C_{N1}}}}$ that is ${{A_N}}$, Bob possess the first particle of ${\left|  +  \right\rangle ^{{B_N},{C_{N2}}}}$ that is ${{B_N}}$. Charlie have ${{C_{N1}}}$ and ${{C_{N2}}}$ at disposal.  These two auxiliary states can be $\frac{1}{{\sqrt 2 }}\left( {\left| {00} \right\rangle  + \left| {11} \right\rangle } \right)$ or $\frac{1}{{\sqrt 2 }}\left( {\left| {01} \right\rangle  + \left| {10} \right\rangle } \right)$, which work the same way. Particles $A$ and $B$ are  respectively strong coupled to its auxiliary system's particles ${A_N}$ and ${B_N}$ by Alice and Bob. The strong couplings can be CNOT gate between the state system particle and one of the auxiliary particles. That is particle $A$ is coupled to one of the particle of $\frac{1}{{\sqrt 2 }}\left( {\left| {00} \right\rangle  + \left| {11} \right\rangle } \right)$ and so do particle $B$. Then, strong measurements are respectively operated by Alice and Bob on  ${A_N}$ and ${B_N}$, these measurement operations convert quantum particles into classical bits and lead a transfer of quantum state from ${\rho _{ABC}}$ to ${\rho _{{C_{N1}}{C_{N2}}C}}$. These measurements can be realized through a projecting base state $\left| 0 \right\rangle $ done by Alice and Bob, and the result of measurements is ${\rho _{ABC}} = {\rho _{{C_{N1}}{C_{N2}}C}}$. If the bases projecting on the auxiliary systems are not $\left| 0 \right\rangle $, then we got the state ${\rho _{{C_{N1}}{C_{N2}}C}}$ which is the local operating result of ${\rho _{ABC}}$, and these would not change the correlations that we obtain at the end of this model.

Charlie continue to deal with ${\rho _{{C_{N1}}{C_{N2}}C}}$ at his disposal. At first, he prepare three auxiliary systems to make ${C_{N1}}$, ${C_{N2}}$, and $C$ to be coupled to. The auxiliary systems ${\left|  +  \right\rangle ^{{{C'}_{N1}}{{C''}_{N1}}}}$, ${\left|  +  \right\rangle ^{{{C'}_{N2}}{{C''}_{N2}}}}$ and ${\left|  +  \right\rangle ^{C'C''}}$ can be also $\frac{1}{{\sqrt 2 }}\left( {\left| {00} \right\rangle  + \left| {11} \right\rangle } \right)$ or $\frac{1}{{\sqrt 2 }}\left( {\left| {01} \right\rangle  + \left| {10} \right\rangle } \right)$ which are shown in Fig. \ref{FIG2}.. A base state $\left| 0 \right\rangle $ is projected on one of the two particles of ${\left|  +  \right\rangle ^{{{C'}_{N1}}{{C''}_{N1}}}}$ to complete measurement on ${{{C''}_{N1}}}$, this process gives the result that ${{{C'}_{N1}}}$ is in the quantum state that ${C_{N1}}$ is in. The similar process happened with ${C_{N2}}$ and $C$, and ${{C'}_{N2}}$ and ${C'}$ respectively take the state of each of them. The measurement base state in this process is allowed to be not $\left| 0 \right\rangle $, which would not lead any influence on the correlations we aim to obtain on condition that the three measurement bases are in the same state. In the next job, Charlie arrange weak measurements on the four line which spread out into from ${\rho _{{C_{N1}}{C_{N2}}C}}$ in Fig. \ref{FIG2}. The weak measurement operators ${A_{ji}}$ in Fig. \ref{FIG3} are laid out in  Table \ref{table1}.  ${\rho _{ji}}$ is the point devices that we can read average position shift $\delta {q_{ji}}$ and average momentum shift $\delta {p_{ji}}$ from, and we can calculate weak value according these shifts which are caused by a postselection simultaneously.
\[{\rm{Re}}\left( {{W_{ji}}} \right) = \frac{1}{\mathtt g}\delta {q_{ji}},\;\;\;{\rm{Im}}\left( {{W_{ji}}} \right) = \frac{1}{\mathtt g}\delta {p_{ji}}\]

Eq. \ref{eq4} gives the necessity of taking into account each element of the baseset acts as postselected state, and this baseset is mutually unbiased to the baseset of the system whose correlation is that we want to get. All postselected states are laid out in Table \ref{table2}, the number of which is eight determine the times of experiment.

\section{CONCLUSIONS AND DISCUSSIONS}
\label{sec:4}

We have presented a model which can be used to obtain correlations of  unknown quantum states. Weak measurements and auxiliary systems were incorporated in this model and the  information we need to know in advance are the dimensions and baseset of the quantum states taken. An example has been given to illustrate this model. The model in this paper can extend to more particles circumstances, and the number of auxiliary system increase with the number of state particles. It can also extend to high-dimensional circumstances on the precondition that the mutually unbiased bases of the space can be found \cite{ijq-8-535,pra-86-022311}, and the more dimensions is the more weak measurements and postselections we need to make. So the generalization of this model to the multipartite system or high-dimensional circumstances is another challenging issue and needs further considerations.

\begin{acknowledgments}
We gratefully acknowledge conversations with Jianlian Cui who is the teacher of Department of Mathematical Sciences, Tsinghua University. This work was supported by the National Natural Science
Foundation of China (Grants No. 11505125).
\end{acknowledgments}

\end{document}